\documentclass[aps,prd,preprint,superscriptaddress,showpacs,showkeys]{revtex4}
\usepackage{amssymb,amsmath}
\usepackage{graphicx} 
\usepackage{dcolumn}  
\usepackage{epsfig}   
\usepackage{pstricks}
\begin{document}


\title{On Quantum Fields at High Temperatures}


%
%

\thanks{\emph{Based on the talk presented by one of us (T.G.) at the 7th International Conference on New Frontiers
in Physics, Kolymbari, 4-12 July 2018, Crete.}}
\author{I. Bischer}
\affiliation{Max-Planck-Institut f\"ur Kernphysik, Postfach 103980, D-69029 Heidelberg, Germany}
\email[]{bischer@mpi-hd.mpg.de}
\author{T. Grandou}
\affiliation{Universit\'{e} de Nice-Sophia Antipolis,\\ Institut Non Lin\'{e}aire de Nice, UMR CNRS 7335; 1361 routes des Lucioles, 06560 Valbonne, France}
\email[]{Thierry.Grandou@inln.cnrs.fr}
\author{R. Hofmann}
\affiliation{Institut f\"ur Theoretische Physik, Universit\"at Heidelberg, Philosophenweg 16, 69120 Heidelberg, Germany}
\email[]{hofmann@thphys.uni-heidelberg.de}


\date{\today}

\begin{abstract} Revisiting the fast fermion damping rate calculation in a thermalized QED and/or QCD plasma at 4-loop order, focus is put on a peculiar perturbative structure which has no equivalent at zero-temperature. Not surprisingly and in agreement with previous $C^\star$-algebraic analyses, this structure renders the use of thermal perturbation theory quite questionable.
\end{abstract}


\maketitle


\section{\label{SEC:1}Introduction}

Let us start by quoting R.D. Pisarski on damping rates in hot gauge theories \cite{Rob}: 
``It is really surprising how difficult it is to calculate damping rates in hot gauge theories"
Even though a so-called {\textit{Resummation Program of Hard Thermal Loops}} (leading-order-in-the-coupling thermal fluctuations at high temperature) has been devised to obtain gauge-invariant, complete results on damping rates and other dynamical observables (and has succeeded to some extent \cite{RoBraat, RoBraat1,FreTayl}), serious difficulties have constantly bounced back and forth due to the {\textit{infrared sector}}. 

Literature testifies of two major obstructions to the Resummation Program that have become textbook material \cite{michel}. The first of these was discovered around the same time as the Resummation Program itself \cite{1st} in applying it to damping rates. The second one was discovered a few years later while using the Resummation Program to evaluate the soft photon emission rate out of a thermal Quark-Gluon Plasma \cite{2nd}. In both situations, the hot gauge theories' infrared sectors were recognized to be at the origin of two singular results. In the latter case, however, a thorough analysis has proven that the alluded singularity is an incorrect one, and that the soft photon emission rate, as determined by the Resummation Program, comes out regular when properly evaluated \cite{tg, tg1,kbtg}. In the former case, our present revisiation of the problem points out that the singular result has been derived unduly. As it turns out, infrared intricacies have long masked a more fundamental difficulty which appears to be inherent to the perturbative approach itself.
\par
The long known and crucial issue concerning interacting covariant quantum field theories at high temperature, is to know wether or not they can admit perturbative treatments. On the basis of a concrete $4$-loop damping rate calculation, the current paper aims at disclosing this fundamental difficulty, once infrared singularities cancellations are established. What remains in effect is some \emph{infrared enhancement mechanism} which prevents any order of a perturbative expansion in the coupling constant to be \emph{complete}, the zeroth order one included. For combinatorial reasons, a diagrammatic control of this order-by-order incompleteness appears out of reach, a situation which pleads in favour of non-perturbative methods.

\section{History}

\subsection{A warning from $C^\star$-algebras}

Aside from the older \emph{Matsubara} imaginary time formalism, well adapted to the calculations of thermodynamical quantities \cite{Matsubara}, the first \emph{real-time} formalism accounting for a finite thermodynamical temperature $T$ (and/or chemical potential $\mu$) appeared in 1974 and was based on the \emph{Dolan-Jackiw} propagator \cite{1974},
\begin{equation}\label{74}
D(P)=\frac{i}{P^2-m^2+i\varepsilon}+2\pi n_B(p_0)\,\delta(P^2-m^2)
\end{equation}where $p_0$ is the energy component of the 4-vector $P$, and $n_B$ is the ordinary \emph{Bose-Einstein} statistical distribution, $n_B(p_0)=\frac{1}{e^{|p_0|/T}-1}$. The perturbation theory based on (\ref{74}) though couldn't avoid ill defined products of singular distributions, like $\bigl[\delta(P^2-m^2)\bigr]^N$, and it is only through $C^\star$-algebra analyses that the necessity of doubling the number of degrees of freedom was recognised to avoid these ill defined terms \cite{michel}. In this way, well-behaved perturbative expansions could be \emph{formally} devised at non-zero $T$ and/or $\mu$.\par\medskip
The same $C^\star$-algebra analyses, however, were also able to point out serious difficulties. In effect, nothing in the Hilbert representation space obtained out of the \emph{Gelfand-Naimark-Segal} ($GNS$) construction could serve the purpose of defining any reliable perturbation theory; at least in the standard $T=0$ sense \cite{Landsman}. For the cases of sufficiently simple scalar-field-theory examples, alternatives could be proposed. However, their 
practical use is limited, unfortunately \cite{Landsman}.

\subsection{At about the same time though }
\label{sec-2}
Around the same time Braaten-Pisarski and Frenkel-Taylor/Taylor-Wong were able to construct the Resummation Program of Hard Thermal Loops ($HTL$) which is the \emph{effective perturbation theory} ruling the leading order thermal fluctuations at momentum scale $eT$ ($QED$), $gT$ ($QCD$). This was   \emph{necessary} because for momenta on the order of ${\cal{O}}(k)=eT$ one-loop corrected propagators are on the same order of magnitude as bare ones. For example in a six dimensional scalar field theory with cubic self interaction $g\varphi^3_6$ one has (with $K^2=k_0^2-k^2$, $Q_0$ a Legendre function and $C^{st}$ a numerical constant)
\begin{equation}
 Re\,\Sigma_{RR}^{HTL}(k_0,k)\simeq C^{st}\,{[\frac{g^2T^2}{k^2}]}\,{K^2}\ \frac{k_0}{k}\ln\frac{k_0+k}{k_0+k}=C^{st}\,{[\frac{g^2T^2}{k^2}]}\,{K^2}\ 2\frac{k_0}{k}Q_0(k_0/k)=\mathcal{O}({K^2})\,,
  \end{equation}so that propagators must be `re-summed'. In a $R/A$ real-time formalism (with $A$dvanced and $R$etarded bare propagators \cite{michel}), one will accordingly define the (retarded) resummed propagator as \begin{equation}
  {}^*D_{RR}(K)=\frac{i}{K^2-\Sigma^{HTL}_{RR}(k_0,k)+i\varepsilon k_0}\,.\end{equation}
  The resummation program has been successful in solving the so-called \emph{gluon damping rate puzzle} \cite{RoBraat} and enjoys remarkable properties. The $HTL$ are $1$-loop order $e^2T(g^2T)$ gauge invariant quantities and obey Ward identities, even in $QCD$. This sort of \emph{abelianisation} of $QCD$ in the high temperature limit can be made one step more explicit by observing that the effective action for $HTL$ in $QCD$ only differs from the one of $QED$ by the \emph{Lie-algebra valuation} of the gauge fields, $A_\mu\rightarrow \sum A^a_\mu T^a$. This observation, which is obvious in the fermionic sector, is more tricky in the bosonic one, but an effective action for $HTL$s like,
  \begin{eqnarray}&&{{m_i}^2\over 2}\int_0^\infty{\rm{d}}\lambda\int_{-\infty}^{+\infty}{{\rm{d}}\sigma\over 2{\sqrt{\pi}}} \ e^{-\sigma^2\over 4}\ \biggl\langle {\widehat K}^\alpha{\widehat K}^\beta\,{\it{Tr}}\biggl\lbrace \bigl[D_\mu,D_\alpha\bigr]\ e^{-\sigma{\sqrt{\lambda}}\  {\widehat K}\cdot D}\nonumber\\&& \times\,\bigl[D^\mu,D_\beta\bigr]e^{+ig\sigma{\sqrt{\lambda}}\ {\widehat K}\cdot A}\biggr\rbrace \biggr\rangle\,,\ \ \ \ \ \ {m_\gamma}^2=\frac{{e^2T^2}}{ 6}\,, \ \ \ {m_g}^2=C_A\frac{{g^2T^2}}{6} +C_F\frac{{g^2T^2}}{12}\,,\end{eqnarray}
where the large brackets stand for a 3-dimensional spatial angular average (${\widehat K}=(1,{\hat k}),\ K^2=0$), encompasses both ${\cal{L}}^{QED}_\gamma$ and ${\cal{L}}^{QCD}_g$, differing only in {{the Lie-algebra valuation of $A_\mu$}} \cite{tgmpla}.

\subsection{Infrared problems or anything else?}
Apart from a \emph{first} series of \emph{proper infrared} ($IR$) and \emph{mass} singularities, which were all revealed to be erroneous, it became textbook material that the resummation program apparently meets two serious $IR$ obstructions, as alluded to in the Introduction. These are,\par\medskip
1. The logarithmic divergence of a rapid ($v\rightarrow 1$) massive fermion damping rate moving through a plasma on the fermion's mass shell at high temperature $T$, Fig.1. With $1+n_B(k_0){\simeq} {T\over k_0}+{1\over 2}+..$ one obtains 
\begin{equation}\label{ms}
\gamma(E,p){\simeq}\lim_{{v}=1}\frac{e^2T}{2\pi}\int^{k^\star}_{E|1-{v}|/2}\,\frac{{\rm{d}}k}{k}\,+\, regular.
\end{equation}Kinematics in this case can be selected so as to make the mass shell and high velocity limits one and the same limit ($v=1$).
\begin{figure}
\centering
{\includegraphics[width=0.58\textwidth]{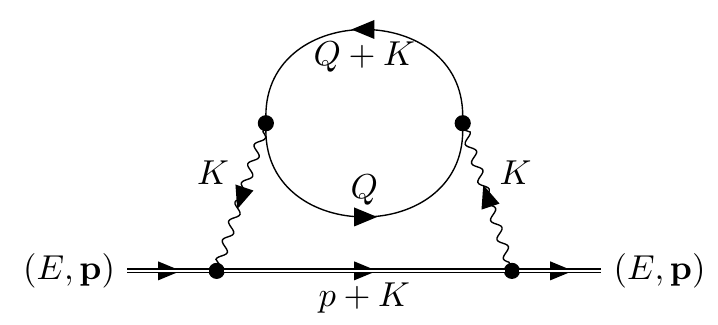}}\
\caption{Basic diagram for the rapid fermion damping rate.}
\label{fig-1}
\end{figure}
\par\medskip
2. The collinear singularity of the soft photon emission rate out of a quark-gluon plasma: $Im\,\Pi_R(Q)$ proportional to ($D=4+2\varepsilon$)
\begin{equation}\label{sing.}
{{\frac{C^{st}}{\varepsilon }}}\!\int {\frac{{d^{4}P}}{(2\pi )^{4}}}\ \delta ({%
\widehat{Q}}\!\cdot \!P)\ (1-2n_{F}(p_{0}))\sum_{s=\pm 1,V=P,P^{\prime }}\pi
(1-s{\frac{v_{0}}{v}})\beta _{s}(V)\,, 
\end{equation}
where $\beta_{s}(V) $ is the space-like part of the fermionic \textit{spectral densities} and 
$V=P,P'$ with $P'=P+Q$ \cite{michel}. 
\par
However, while divergence 1. does not exist, being an ill-posed problem (see below), divergence 2. does not exist either \cite{kbtg}. The singular result (\ref{sing.}), in fact, is due to the following double entwined angular integral which, to our knowledge, is impossible to compute numerically:
\begin{equation}\label{W}
W(P,P^{\prime })=\sum_{s,s'=\pm}\int {\frac{\mathrm{d}{\widehat{K}}}{4\pi }}\int {%
\frac{\mathrm{d}{\widehat{K}^{\prime }}}{4\pi }}\ {\widehat{K}}\!\cdot \!{%
\widehat{K}^{\prime }}{\frac{{\widehat{K}}\!\cdot \!{\widehat{P}}_{s}\ {%
\widehat{K}^{\prime }}\!\cdot \!{\widehat{P}^{\prime }}_{s^{\prime }}+{%
\widehat{K}}\!\cdot \!{\widehat{P}^{\prime }}_{s^{\prime }}\ {\widehat{K}%
^{\prime }}\!\cdot \!{\widehat{P}}_{s}-{\widehat{K}}\!\cdot \!{\widehat{K}%
^{\prime }}{\widehat{P}}_{s}\!\cdot \!{\widehat{P}^{\prime }}_{s^{\prime }}}{%
({\widehat{K}}\!\cdot \!P+i\epsilon )({\widehat{K}}\!\cdot \!P^{\prime
}+i\epsilon )({\widehat{K}^{\prime }}\!\cdot \!P+i\epsilon )({\widehat{K}%
^{\prime }}\!\cdot \!P^{\prime }+i\epsilon )}}\,.
\end{equation}
An \emph{estimated} singular behaviour of (\ref{W}) was retained, giving rise to (\ref{sing.}), whereas a full, exact (cross-checked) calculation of $W(P,P^{\prime })$ displays a series of mass singularities of strengths $\varepsilon^{-1}$ and $\varepsilon^{-2}$ which cancel exactly out among themselves, leaving a regular result \cite{kbtg}. Under the disguise of {\emph{IR}} problems, what shows up instead is a {\emph{structural obstruction}} to perturbative attempts, not perceived \emph{as such} by the $C^\star$-algebra analysis because it is disclosed in a specific way through higher number of loop calculations. \par
Now in order to see this, it is appropriate to continue a short while with \emph{history}.

\subsection{Back to \emph{History}}

In $QCD$, a similar effective perturbation theory rules softer, order ${g^2T}$ thermal fluctuations \cite{g2T} and enjoys the same remarkable properties as the $HTL$-Resummation Program ruling the leading order thermal fluctuations at momentum scale $gT$.

 For external momenta $k_0\leq k\sim g^2T$ the so-called {\emph{ultra-soft} amplitudes} are {{as large as the corresponding $HTL$ and tree-level ones}}. Like the $HTL$ ones, ultra-soft amplitudes are gauge-fixing independent (linear gauge conditions like in covariant and Coulomb-like gauges) and satisfy simple Ward identities. Differences exist however. (i) Ultra-soft amplitudes have no abelian counterpart (QED has no $HTL$-amplitudes other than those corresponding to diagrams with external N-photons and 2-electrons). (ii) And a most noticeable difference with soft amplitudes is that ultra-soft amplitudes receive contributions from an infinite series of multi-loop diagrams (\emph{ladder diagrams}) while $HTL$ amplitudes are one-loop diagrams only. \par
 It is also interesting to remark that this new perturbation theory becomes effective at such a momentum scale which allows it to be derived out of \emph{kinetic theory} and out of a generalisation of the \emph{Vlasov equation} for ordinary plasmas \cite{g2T, g2T1}.

\section{Higher order fluctuations}
\subsection{Three-loop order contributions to the polarisation tensor.}
\begin{figure}
\centering
{\includegraphics[width=0.48\textwidth]{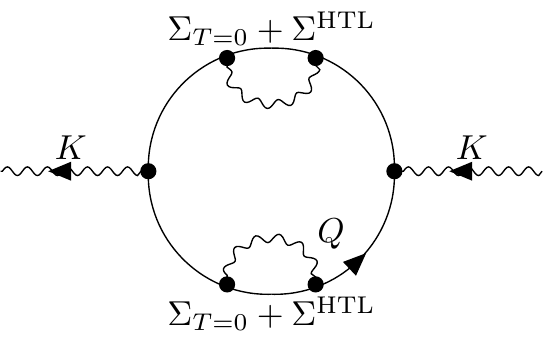}}\
\caption{A 3-loop contribution to the polarisation tensor $\Pi(K)$ which is gauge-fixing independent with $\Sigma^{HTL}$ insertions only.}
\label{fig-2} 
\end{figure}
Higher order fluctuations are to be analysed by means of diagrams as one can no longer rely on \emph{semi-classical} guide lines \cite{ITR}. Such a diagram as that of figure~\ref{fig-2} does not seem to define any \emph{gauge invariant} fluctuation at a momentum scale other than the {ultra-soft} scales, ${e^2T}$ ({{$QED$}}) and ${g^2T}$ ($QCD$). In the small $k/T$ limit, in effect, this fluctuation contributes to the \emph{longitudinal} piece of  $\Pi(K)$ an amount,
\begin{equation}
\delta\Pi^{(3)}_{L}(k_0,k)\simeq K^2\,\,{[\frac{e^2T}{k}]^3}\, \,\delta F^{(3)}_{L}({k_0, k},T)\,,
\end{equation}
where $\delta F^{(3)}_{L}$ (some explicit function of $k_0,k,T$) generalizes in this 3-loop order calculation the previous 1-loop case of Legendre functions $Q_0(x)$ and $Q_1(x)=xQ_0(x)-1$, 
\begin{equation}\mathcal{O}(\delta F_L^{(3)})_{k_\mu=\mathcal{O}(e^2T)}=\mathcal{O}(Q_0(k_0/k), Q_1(k_0/k))=\mathcal{O}(1)\,.\end{equation}
However, this 3-loop fluctuation in the polarisation tensor remains \emph{stuck} at the ultra-soft scale $e^2T$, just providing an example of a higher number of loops diagram contributing to the ultra soft amplitudes.
\par\medskip
Now, the 1-loop $T=0$ renormalised {\textit{e.m.}} vertex (external lines on mass shell $Q^2=(Q+K)^2=m^2$) reads ($QED$),
\begin{equation}\label{em}
 \Gamma^{\mu;ren.}_{T=0}(K)= \gamma^\mu\,F^{ren.}_1(K^2)+\frac{i}{2m}\,\sigma^{\mu\nu}k_\nu\,F^{ren.}_2(K^2)\,,
\end{equation}
and remarkably, $F^{ren.}_1$ and $F^{ren.}_2$ are gauge-invariant at any order of perturbation theory. At $\sinh^2(\theta/2)=-K^2/4m^2$ ($K^2$ is kinematically constrained to be negative), where $m$ stands for the electron (or quark) mass, $F_1$ is given as \cite{IZ},
\begin{equation}\label{F1}
F_1(K^2)=-{\alpha\over \pi}\biggl[(1+\ln{{\mu}\over m})(1-\theta\coth\theta)+2\coth\theta\int_0^{\theta/2}\mathrm{d}\varphi\,\varphi\tanh\varphi+{\theta\over 4}\tanh{\theta\over 2}\biggr]\,,\end{equation}
with the small $\theta$-limit being,
    \begin{equation}\label{F11}F_1(K^2)\simeq {\alpha\over 3\pi}{K^2\over m_e^2}\,(\ln{m\over \mu}-{3\over 8})\ \rightarrow\ -{\alpha\over 8\pi}{K^2\over m_e^2}\,,\end{equation}
while $F_2$ is irrelevant at this order. In (\ref{F1}) and (\ref{F11}) the mass $\mu$ regulates an $IR$ singular behaviour compensated for by another \emph{cut} of the same diagram, both contributing to the non-zero imaginary part of $\delta\Pi^{(3;ren.)}_L(k_0,k)$, while the \emph{intermediate singularity} $\ln{m\over \mu}$ cancels out, in agreement with the $T=0$ context of the $KLN$ theorem \cite{Muta}. 

The $3$-loop fluctuation of the internal photonic $K$-line, depicted in figure~3,
 is made out of pieces which are \emph{gauge invariant} separately and should give rise to a gauge invariant result by construction.  This can also be checked by an explicit calculation giving \cite{ITR}, 
\begin{equation}
{K^\mu\,\Pi^{(3;ren.)}_{\mu\nu}(k_0,k)=0}\,,\end{equation} and in the small $k/T$-limit, one obtains this time, with $C$ a calculable constant,
\begin{equation}\label{good}
\delta\Pi^{(3;ren.)}_L(k_0,k)=-K^2\,{C\over 24\pi^2}{\bigl[{e^3T\over k}\bigr]^2}\,\delta \mathcal{F}^{(3;ren.)}_L(k_0,k)\,.
\end{equation}
That is, $T=0$ renormalised, the contribution $\delta\Pi^{(3;ren.)}_L(k_0,k)$ therefore identifies with  $K_\mu\sim{e^3T}$ a newly emergent momentum scale of {gauge invariant} \emph{vacuum and statistical} mixed fluctuations: A re-summation is in order along the $K$-line whenever $K$ is on the order of this momentum scale. 
\begin{figure}
\centering
{\includegraphics[width=0.58\textwidth]{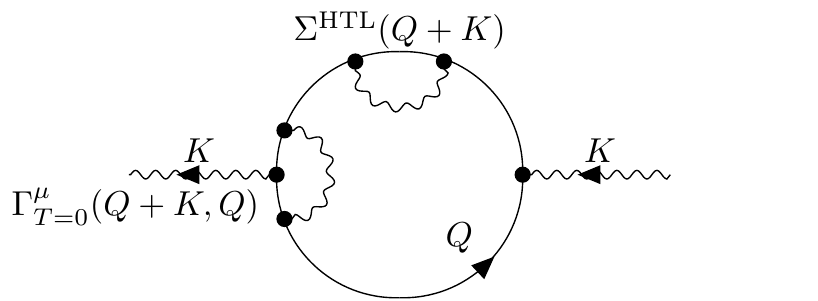}}\
\caption{A gauge-fixing independent $3$-loop contribution to $\Pi(K)$.}
\label{fig-3}
\end{figure}
In the rapid fermion damping rate calculation though, \emph{transverse} 
contributions contribute in addition to the longitudinal ones as,  
\begin{equation}\label{stargamma}{}\gamma (E, p) =\lim_{v=1}{e^2\over 2\pi v}\int k^2{\rm
d}k\int_{-v}^{+v}{{\rm d}x\over 2\pi}\ (1+n_B(kx))\biggl\lbrace
{}\rho_L
(kx,k)+(v^2-x^2){}\rho_T(kx,k)\biggr\rbrace\,.\end{equation}
In the $R/A$-real time formalism (${\eta\rightarrow 0^+}$), one has
\begin{equation}\label{spectral1}{{}\rho_{L}}_{R/A}(k_0,k)=2\ {\mathcal{I}}m\ {K^2\over k^2}\ {1\over K^2-\Pi_{L}(k_0\pm
i\eta,k)}\,,\ \ \ \ \ {{}\rho_{T}}
_{R/A} (k_0,k)=2\ {\cal{I}}m\ {1\over
K^2-\Pi_{T}(k_0\pm i\eta,k)}\,,\end{equation}
where the two $R/A$-{spectral densities} ${{}\rho_{L}}$ and ${{}\rho_{T}}$ have non-trivial parts (\textit{i.e.}, order $e^{2n}$ in an order $e^{2n}$ calculation of the polarisation tensor) if and only if the imaginary parts of $\Pi_{L,T}(k_0,k)$ are non-zero. In the case of the re-summation program, for example, this condition is met thanks to the Legendre functions $Q_0$ and $Q_1$ which develop imaginary parts at space-like momenta, $K^2\leq 0$,
\begin{equation}\label{PiHTL}\Pi_L\rightarrow \Pi_L^{(HTL)}
=-2m^2{K^2\over k^2}+m^2{K^2\over k^2}{k_0\over k}\ln{k_0+k\over
k_0-k},\ \ {\rm{and}} \ \ \Pi^{(HTL)}_T=m^2-\Pi^{(HTL)}_L/2\,\end{equation}with order $e^2T^2$ ($QED$) and order $g^2T^2$ ($QCD$) \emph{thermal masses} squared, $m^2$.
This condition is met also in the case of the $3$-loop example of (\ref{good}) for which, with obvious notations, one has,
\begin{eqnarray}\label{F3}
\delta \mathcal{F}^{(3;ren.)}_L(k_0,k)={1\over \pi} \bigl[{k_0\over k}Q_0({k_0\over k})+Q_1({k_o\over k})\bigr]\left(\int_0^\infty{\mathrm{d}x\over x} {\mathrm{d}\over \mathrm{d}x}\ x\tanh{x\over 2}\right)^{ren.}\\-{1\over 4\pi^2}Q_1({k_o\over k})\left(\int_0^\infty{\mathrm{d}x\over x}\tanh{x}\right)^{ren.}\nonumber\\ +{1\over 8\pi^2}\biggl[{k\over k_0}\!\int_0^\infty\!{\mathrm{d}x }\!\! \int_0^x{\mathrm{d}x_0\over x_0-x } [{\tanh{x_0\over 2}\over x_0}\ln{k_0x_0+kx\over k_0x_0-kx}-{\tanh{x\over 2}\over x}\ln{k_0+k\over k_0-k}]\biggr]^{ren.}\,.\end{eqnarray}
Getting back to the transverse contributions, one finds that in the small $k/T$-limit they display 
a leading order part of (same diagram), 
\begin{equation}\delta\Pi_T^{(3;ren.)}=
-C\frac{\bigl[e^3T\bigr]^2}{24\pi}\,\bigl[{T\over k}\bigr]^2\left(\int_0^\infty \frac{{\mathrm{d}x}}{x}\int_{0}^{x}\mathrm{d}y\,y\tanh{xy\over 2}\right)^{ren.}\,.\end{equation}
This implies the following orders of magnitude for the longitudinal and transverse spectral densities 
\par\medskip
\begin{equation}
\mathcal{O}\left({}^{(3)}\rho_T(k_0,k)\right)={\mathcal{O}\left({k^2\over T^2}\right)^2}\times \mathcal{O}\left({}^{(3)}\rho_L(k_0,k)\right)\,.\end{equation}
Accordingly, transverse-degrees contributions preserve the peculiarity of the longitudinal degree contribution (\ref{good}) and furthermore need no re-summation. The same analysis carried out in the case of two-loop contributions to $\Pi^{(2;ren.)}_{L,T}(k_0,k)$, corresponding to the diagrams of Figures 4 and 5, after the detailed balance of all infrared singularity cancellations has been checked \cite{ITR}, allows one to express the rapid fermion damping rate contributions as follows
\begin{figure}
\centering
{\includegraphics[width=0.73\textwidth]{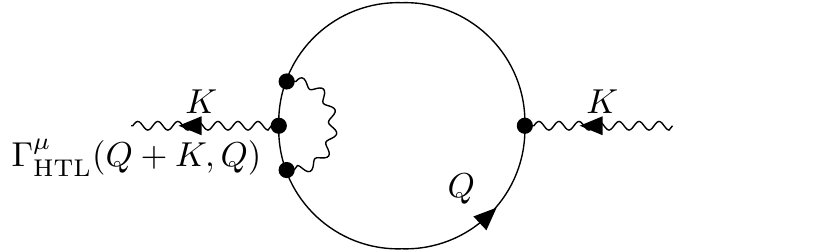}}\
\caption{An gauge-fixing independent $2$-loop contribution to $\Pi(K)$.}
\label{fig-4}
\end{figure}
\begin{figure}
\centering
{\includegraphics[width=0.48\textwidth]{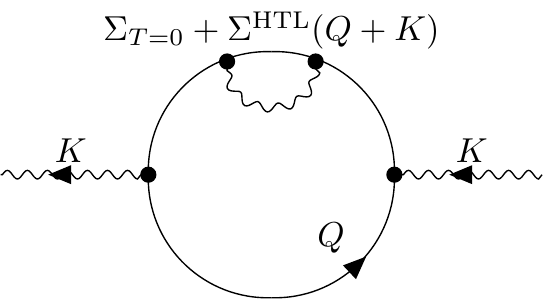}}\
\caption{A $2$-loop contribution to $\Pi(K)$. Gauge-fixing independent with $\Sigma^{HTL}(Q+K)$ inserted only.}
\label{fig-5}
\end{figure}
\begin{equation}\label{22}\delta_{{n}}\,\gamma (E,
p)\simeq{e^2T\over 2\pi }\int_{e^{[{n}+{1\over 2}]}T}^{e^{[{n}-{1\over 2}]}T}\,k{\rm d}k\int_{-k}^{+k}{{\rm d}k_0\over
k_0} \bigl\lbrace {}^{({n})}\rho_L
(k_0,k)+{}^{({n})}Transv.\bigr\rbrace\,\end{equation}
with 
\begin{equation}
{}^{({n})}\rho_L
(k_0,k)=\frac{1}{k^2+C^{({n};ren.)}\bigl[e^{{n}}T\bigr]^2\,\delta\mathcal{F}^{({n};ren.)}_L(k_0,k)}\,.\end{equation}
It is worth observing that contrarily to the $T=0$ case, for which one would write
\begin{equation}\label{0TT}\rho^{(3)}_{R,L}(k_0,k)=2\ {\cal{I}}m{K^2\over k^2}\ {1\over K^2-\Pi_L^{(1)}(k_0,k)-\Pi_L^{(2;ren.)}(k_0,k)-\Pi_L^{(3;ren.)}(k_0,k)+i\varepsilon k_0}\,\end{equation}
and obtain subleading order $e^4$ and $e^6$ corrections to ${}^\star\rho_L
(k_0,k)$ and then to $\gamma(E,P)$, one is here in a situation where each scale of invariant fluctuations 
contributes additively and independently. This peculiar \emph{layered structure} has no equivalent at $T=0$. Notice that the $HTL$-layer is at ${n}=1$ with ${}^{({1})}\rho_L\equiv{}^\star\rho_L $, $  \delta\mathcal{F}^{({1})}_L(k_0,k)=Q_1(k_0,k)$ as given in (\ref{PiHTL}) and $C^{(1)}={1/3}$. \par
In (\ref{22}) each gauge-invariant fluctuation is assigned an effective range,
 \begin{equation}\label{usage}{e^{[{n}+{1\over 2}]}T}\leq |k_0|\leq k\leq {e^{[{n}-{1\over 2}]}T}\end{equation}
 in agreement with a common, still somewhat arbitrary usage \cite{michel}. Now, the most interesting point is that independently of the way the validity range of (\ref{usage}) may be redefined, the following result is preserved,
 \begin{equation}\label{01}
\delta_{{n}}\,\gamma (E,
p)={e^2T\over 2\pi }\,\mathcal{O}_{{n}}\left(1\right)\,, \ \ \ {n=1,2,3,..}\,,\end{equation}where the flexibility in the way (\ref{usage}) can be decided gets entirely reflected in the various values the constants $\mathcal{O}_{{n}}\left(1\right)$ may take. This means that even at leading order (order of $e^2T$ in this very case) such an observable as the rapid massive fermion damping rate travelling through a thermalized plasma, receives contributions from a number of invariant fluctuations. Not only \emph{soft}, but \emph{ultra-soft} and even \emph{softer} fluctuations must be taken into account.
\par\medskip
An important by-product of this result is that the famous infrared divergence, plaguing the rapid fermion damping rate by its mass shell, see Eq.(\ref{ms}), arises artificially,
\begin{equation}
\gamma(E,p)\simeq\lim_{v=1}\frac{e^2T}{2\pi}\int^{k^\star}_{E|1-v|/2}\,\frac{{\rm{d}}k}{k}\,,\ \ \ \ \ k^\star=e^{[{1}-{1\over 2}]}T\,.
\end{equation}
Because of the existence of softer invariant fluctuations contributing on the same order of magnitude, in effect, the $HTL$ effective perturbation theory or re-summation program cannot be extended down to $k=0$ but only to some lower limit, such as $k_{min}=e^{[{1}+{1\over 2}]}T$, to comply with (\ref{usage}). As a consequence, there is again no real infrared problem. Definitely, there is something else though which we will discuss in what follows.

\section{Discussion}

This communication summarises a long analysis \cite{ITR} whose more detailed aspects, like at $n=2$ and $n=3$, the detailed balance of $IR$ singularity compensations, and also the relevance of using (\ref{em}) have not been spelled out here. Out of this analysis, the following points appear to be worth retaining:
\begin{itemize}  
\item {The ultra-soft fluctuations at momentum scale ${e^2T}$ are not as {\textit{terminal}} as thought initially. Softer invariant fluctuations do exist, such as those of momentum order ${e^3T}$, emerging out of specific higher number of loop diagrams. }

\item {By construction these fluctuations are gauge invariant as can be checked also by explicit calculations. They are mixed fluctuations in that they result from an interplay of vacuum renormalised fluctuations with statistical/thermal ones, and they require a sufficiently high number of loops in diagrams in order to come about (note that \textit{a priori} an infinite number of similar invariant fluctuations can be constructed and that other invariant fluctuations are not excluded either).}

\item{They contribute to the {{zeroth-order}} approximation in the rapid fermion damping rate $\gamma(E,p)$, and subsequent corrections to $\gamma(E,p)$ will of course be confronted with the same situation because they are induced by the high temperature infrared enhancement. }

\item {Apart from the issue of mastering a tower of invariant fluctuations so as to control \emph{the leading part only} of a given observable and in addition to the fact that none of these equally important fluctuations does enjoy a non-equivocal determination of its contribution to $\gamma(E,p)$ (\textit{i.e.} the $\mathcal{O}_{{n}}\left(1\right)$ of (\ref{01}) depend on the somewhat arbitrarily defined ranges (\ref{usage}))} ..

\item{.. the very principle of range definition separation, {\textit{i.e.}}, the clear-cut separation of momentum scales {$T$, $g(T)T$, $g^2(T)T$, $g^3(T)T$}, \textit{etc..} is deprived of any physical realisation even at very high $T$, as checked up to $10^{25-30}T_c$ in the pure Yang-Mills case \cite{Forcrand}. A by-product is that, as noticed by J.P. Blaizot in 1999, a condition necessary for a consistent implementation of the renormalisation group {\textit{\`a la}} Wilson fails to be met.}

\end{itemize}

 \section{Conclusion}
 Of all this it seems reasonable to conclude that the warning from $C^\star$-algebraic analysis \cite{Landsman} is justified. At high enough temperatures at least, we can see that standard, formally well-defined perturbative expansions are not able to address the issue of \emph{completeness} in a leading order calculation of a basic \emph{dynamical} observable such as a damping rate. \emph{Static} observables are most conveniently evaluated within the Matsubara imaginary time formalism, but as long displayed by the famous \emph{Linde problem}, they are not preserved from this flaw either \cite{Linde}. 
 \par
 High $T$ quantum field theories definitely call for non-perturbative methods \cite{Lowdon}. A new $T\neq 0$ formalism was born very recently \cite{Hugo} which by construction (implementing temperature by \emph{compactifying} a spatial rather than the time direction) could perhaps disentangle the infrared sector from the high temperature limit of covariant (not invariant!) quantum field theories. To our knowledge, however, the new formalism has not yet been explored in this respect.
 \par\medskip
 In the fermion damping rate example that we have dealt with things clearly happen to be as they are because of the so-called \emph{infrared enhancement} mechanism. As long observed in the general context of perturbatively accessed 
 quantum field theories, in effect, infrared singularities are nothing but intermediate quantities, no matter how difficult it may be to check their overall cancellations; and the above alluded infrared enhancement mechanism can consistently be viewed as the \emph{imprint} of these intermediate infrared singularities.
 
Infrared problems often hide a more fundamental aspect. At high temperature infrared singularities are known to be much more severe than at zero temperature because of the Bose-Einstein statistical distribution, $1+n_B(kx)={T\over kx}+{1\over 2}+{\cal O}({kx\over T})$ and because of the explicit breaking of \emph{Lorentz invariance}. When properly dealt with, however, all of the generated infrared and mass singularities cancel out, and what remains eventually is a net enhancement of the infrared sector. The consequence, which is emphasized here, is the impossibility for \emph{bare} and \emph{effective perturbative expansions} to yield \emph{complete} answers. Assuming separated enough momentum scales, the $e^nT$s, a most important and new point of the present analysis is that the high temperature infrared enhancement brings a diagrammatically non-controllable number of gauge invariant fluctuations to the same level of importance in the leading order calculation of a physical quantity. Of course, such a situation sheds light on the lingering issue mentioned in the Introduction, that of the relevance of perturbative attempts in the context of high temperatures quantum field theories.

Seen from the $C^\star$-algebraic point of view, this impossibility is the indication that one is not working within the appropriate representation space \cite{Landsman}.
 \par
 Unfortunately, as stated by the end of Subsection 2.1, alternatives starting from a representation space more relevant to the basic observables of a thermal context do not seem to offer a promising approach either, at least for the sake of practical use. Another attempt could possibly rely on the use of \emph{non-perturbative functional methods} which, so far, haven't been used very much in the non-zero $T$ and/or $\mu$ cases. Cases, where this was done can be found in \cite{BI} and in \cite{HTM}. It is interesting enough to take a look at the result in \cite{HTM}, the Bloch-Nordsieck approximated propagator calculation of an energetic fermion of mass $m$ travelling through a $QED$ thermalised plasma in thermal equilibrium at temperature $T$. One finds,
 \begin{eqnarray}
&&{S'}(\omega, z_0)=i \frac{2m}{\omega} \biggl\lbrace \frac{1}{2} \, e^{-i \omega z_0
-\frac{{{A}}^2}{4 \omega^2} z_{0}^{2}}\nonumber\\&& - e^{-\frac{\omega}{T} -
\frac{{{A}}^2}{4 \omega^2}(z_{0}^{2} - \frac{1}{T^{2}})} \,
\cos{\left(\left[\omega - 2 T (\frac{{{A}}}{2\omega T})^{2}\right]
z_{0}\right)} \biggr\rbrace
\end{eqnarray}
where 
\begin{equation}
{{A}}^{2} =  \frac{4{\alpha}}{3\pi} ({\vec{p}}^{\ 2})^2\left(1+(\frac{2\pi T}{p})^2\right) - \frac{4{\alpha}^{2}}{3\pi}  (\vec{p}^{\ 2})^2
\ln{\left(\frac{\vec{p}^{\ 2}}{m^2}\right)}
\end{equation}and 
\begin{equation}z_0=x_0-y_0,\ \vec{p}=\vec{p}(z_0)=\vec{p}(0)\,e^{-\Gamma z_0},\  \Gamma=\frac{2}{{\sqrt{3\pi}}}\frac{{\alpha}c}{\lambda_c}\left(\frac{k_BT}{mc^2}\right)^2\,.\end{equation}
As a function of time travel, $z_0$, one can observe the full complexity of the energetic fermion propagation in the plasma. A complexity that perturbative attempts, at least, may have taken us to suspect.
\par\medskip
It is worth mentioning also that an alternative to the perturbative loop expansion can be found within the approach developed in \cite{Ralf1, Ralf2}, starting out from an explicit construction of the thermal ground state in terms of gauge-field configurations that cannot be reached by small-coupling expansions.

\acknowledgments{I.B. is supported by the IMPRS-PTFS and enrolled at Heidelberg University. We also thank the CNRS for funding a visiting (DR2) position for R.H. at the 'Insitut de Physique de Nice' in summer of 2016, when parts of this work were carried out, and ITP of Heidelberg University for supporting I.B. during parts of this period.}

\end{document}